\documentclass[journal,twoside,web]{IEEEtran}
\usepackage{cite}
\usepackage{amsmath,amssymb,amsfonts}
\usepackage{algorithmic}
\usepackage{graphicx}
\usepackage{textcomp}
\usepackage{comment}
\usepackage{upgreek} 
\usepackage{bm} 
\usepackage{subfigure}
\usepackage{tabularx,booktabs}
\usepackage{titlesec}
\usepackage{hyperref}
\hypersetup{
  colorlinks=true,
  linkcolor=blue,
  filecolor=magenta,
  urlcolor=cyan,
  pdftitle={Overleaf Example},
  pdfpagemode=FullScreen,
}
\def\BibTeX{{\rm B\kern-.05em{\sc i\kern-.025em b}\kern-.08em
T\kern-.1667em\lower.7ex\hbox{E}\kern-.125emX}}

\begin{document}
\title{Multimodal Higher-Order Brain Networks: A Topological Signal Processing Perspective}
\author{Breno C. Bispo, Stefania Sardellitti, \IEEEmembership{Senior Member, IEEE,} Juliano B. Lima, \IEEEmembership{Senior Member, IEEE,} and Fernando A. N. Santos
\thanks{This work was supported  by the  FIN-RIC Project 
TSP-ARK, financed by Universitas Mercatorum under grant n. 20-FIN/RIC. Additional support was provided by CAPES (88881.311848/2018-01, 88887.899136/2023-00), CNPq (442238/2023-1, 312935/2023-4, 405903/2023-5, 200548/2025-5), and FACEPE (APQ-1226-3.04/22).
Bispo  and Lima are with Dept. of Electronics and Systems, Federal University of Pernambuco, Recife, Brazil. Sardellitti is with the Dept. of Engineering and Sciences, Universitas Mercatorum, Piazza Mattei, Rome, Italy. Santos is with the  Dutch Institute for Emergent Phenomena, KdVI, University of Amsterdam, Amsterdam, The Netherlands. E-mails: {breno.bispo, juliano.lima\}@ufpe.br, stefania.sardellitti@unimercatorum.it, f.a.nobregasantos@uva.nl.}
}
}

\maketitle

\begin{abstract}
Brain connectomics is still largely dominated by pairwise-based models, such as graphs, which cannot represent circulatory or higher-order functional interactions. In this paper, we propose a multimodal framework based on Topological Signal Processing (TSP) that models the brain as a higher-order topological domain and treats functional interactions as discrete vector fields.
We integrate diffusion MRI and resting-state fMRI to learn subject-specific brain cell complexes, where statistically validated structural connectivity defines a sparse scaffold and phase-coupling functional edge signals drive the inference of higher-order interactions (HOIs). Using Hodge-theoretic tools, spectral filtering, and sparse signal representations, our framework disentangles brain connectivity into divergence (source-sink organization), gradient (potential-driven coordination), and curl (circulatory HOIs), enabling the characterization of temporal dynamics through the lens of discrete vector calculus.
Across 100 healthy young adults from Human Connectome Project, node-based HOIs are highly individualized, yet robust mesoscale structure emerges under functional-system aggregation. We identify a distributed default mode network-centered gradient backbone and limbic-centered rotational flows; divergence polarization and curl profiles defining circulation regimes with insightful occupancy and dwell-time statistics.
These topological signatures yield significant brain-behavior associations, 
revealing a relevant higher-order organization intrinsic to edge-based models.
By making divergence, circulation, and recurrent mesoscale coordination directly measurable, this work enables a principled and interpretable topological phenotyping of brain function.

\end{abstract}


\vspace{-0.1cm}
\section{Introduction}
\label{sec:introduction}

\IEEEPARstart{A}{dvances} in magnetic resonance imaging (MRI) have enabled non-invasive, whole-brain measurements at millimetric resolution, establishing connectomics as a central framework for studying the brain as a distributed system in which cognition and behavior emerge from coordinated interactions across cortical and subcortical areas \cite{sporns_2009,bassett}. In network neuroscience, functional MRI (fMRI) and diffusion MRI (dMRI) are widely used to estimate functional and structural connectomes, typically modeled as graphs with brain regions of interest (ROIs) as nodes and pairwise functional or anatomical connections as edges \cite{bassett}. This graph-based paradigm has provided key insights into segregation--integration balance, hub architecture, and dysconnectivity patterns \cite{sporns_2013,van2013network,sporns_2019}.

Although graphs are typically used to represent pairwise relations, multivariate functions defined on them may capture coordinated activity across groups of nodes \cite{peixoto2026}. This is especially relevant for fMRI, whose signals are often approximately Gaussian \cite{fmri_gaussianity}; in this regime, higher-order statistics reduce to functions of pairwise covariance \cite{rosas}. Accordingly, linear (pairwise) correlation explains much of functional connectivity and a substantial part of higher-order behaviors observed in fMRI \cite{fmri_gaussianity,rikkert,bispo2026}. Still, an explicit topological domain with physically interpretable operators is of great interest, as it allows edge-centric dynamics to be treated as discrete vector fields and enables the analysis of circulations and general signals on higher-order structures not definable from edges alone \cite{barb_Mag_2020}.
This motivates higher-order representations, such as simplicial (or cell) complexes, which explicitly encode non-dyadic interactions, capture clique/cavity organization, recurrent motifs, and closed-loop coordination in brain networks \cite{Sizemore2018CliquesCavities, rieck2020,billings2021simplicial,anand, bispo}. Such features have improved task decoding, individual identification, and behavioral associations, while time-resolved higher-order analyses increase sensitivity to dynamic functional organization \cite{santoro2024higher,teng}.

Recent works show that functional coordination is dynamic and structurally constrained. Dynamic functional connectivity and edge-centric approaches capture temporal variability through edge time series derived from co-fluctuations or phase-coherence \cite{Faskowitz2022,Zamani2022,Cabral2017,leida}, but remain largely dyadic and incapable to represent higher-order coordination. In parallel, structural connectivity constrains, but not fully determine, functional organization, as anatomical scaffolds account for only part of functional interactions \cite{Suarez2020LinkingSF,Fotiadis2024SFCreview}. These observations motivate multimodal approaches that preserve the structural backbone while allowing functional flow dynamics to express emergent multi-way coordination, including circulatory patterns accessible only through higher-order topological analysis.

In this paper, we introduce a novel perspective for brain analysis based on Topological Signal Processing (TSP)~\cite{barb_2020,barb_Mag_2020}, a powerful framework to analyze higher-order interactions (HOIs) in brain networks, by characterizing functional coordination across edges and higher-order structures. TSP extends Graph Signal Processing (GSP)~\cite{shuman2013} to higher-order topological domains and provides Hodge-theoretic operators and spectral tools for signals defined on simplicial and, more generally, cell complexes~\cite{tsp_cell}. In particular, cell complexes represent relations among groups of nodes through cells of different orders (nodes, edges, and polygonal faces), providing the explicit higher-order scaffold needed to analyze edge-centric dynamics beyond pairwise relations. Within this framework, edge signals are interpreted as discrete vector fields and decomposed into orthogonal gradient, solenoidal, and harmonic components, yielding a physically interpretable description of source-sink and cyclic coordination patterns. While Hodge-based descriptors have been used to characterize cycle/flow structure in thresholded functional connectivity networks~\cite{anand}, a unified TSP pipeline that combines brain topology learning with dynamical analysis on subject-specific cell complexes remains unexplored.

To fill this gap, we model each subject's brain as a second-order cell complex that integrates dMRI and resting-state fMRI (rs-fMRI), thereby combining structural and functional connectivity information. DMRI defines a sparse admissible edge scaffold, while HOIs are inferred from time-resolved functional edge dynamics, yielding subject-specific polygonal faces (triangles, quadrilaterals, and pentagons). This supports multiscale analyses from node-level motifs to mesoscale organization across canonical subsystems and their combinations. We then leverage divergence and curl to characterize (i)~mesoscale source-sink 
polarization, (ii)~higher-order cyclic coordination regimes, and (iii)~their associations with inter-individual variability in cognition and behavior in the Human Connectome Project (HCP) 
cohort~\cite{hcp_dataset,Barch2013}.

Our main contributions are as follows. 
First, we extend~\cite{bispo} from simplicial to cell complexes, enabling richer polygonal HOI representations in a multimodal framework, where statistically validated dMRI connectivity defines the structural scaffold and rs-fMRI edge dynamics drive subject-specific higher-order cell (HOC) inference. Unlike pairwise connectomics, this yields an explicit topological domain in which functional interactions can be represented as edge flows and circulations over HOIs.
Second, we show that higher-order brain topology is strongly scale-dependent, where exact node-level HOIs are highly individualized across subjects, whereas robust and reproducible mesoscale organization emerges when HOIs are aggregated by functional-system combinations. 
Third, we show that the proposed perspective turns brain connectivity into a higher-order topological domain endowed with well known physically interpretable operators, overcoming limitations of graph-based models and other higher-order signal processing approaches~\cite{breno}. Using TSP concepts, we identify significant gradient and rotational regimes and quantify divergence- and curl-based mesoscale patterns, making edge flows, circulations, and recurrent coordination directly measurable and behaviorally relevant.

The paper is organized as follows. Sec.~\ref{sec:methods} reviews the TSP framework and topology learning strategy, Sec.~\ref{sec:cell_learning} presents the proposed multimodal pipeline for inferring subject-specific 2-cell complexes from HCP dMRI and rs-fMRI data, Sec.~\ref{sec:results} reports the empirical findings, and Sec.~\ref{sec:conclusion} concludes the paper.

\vspace{-0.1cm}
\section{Topological Signal Processing Framework}\label{sec:methods}

In this section, we briefly review the main notions of the TSP framework over cell complexes ~\cite{tsp_cell}, then introduce a TSP-based approach to jointly learn the brain topology and sparse signal representations from data.
\vspace{-0.2cm}
\subsection{Topological signal processing background}
\label{subsec:tsp_overview}
The need to extract meaningful information from data has motivated structured representations that capture relational dependencies, such as  topological domains. In this context, GSP \cite{shuman2013} analyzes signals on graph nodes, naturally modeling pairwise relations. 
However, graphs do not support notions of flow circulation or higher-order conservation laws. In particular, operators like curl, which quantify cyclic coordination, require higher-order topological structures.
This is relevant for systems such as brain networks, where coordination often extends beyond simple pairwise relations. Even in fMRI, where pairwise interactions often dominate~\cite{fmri_gaussianity, rikkert, bispo2026}, higher-order representations remain essential to capture the geometry of interactions and define operators that characterize flow organization.
To address this limitation, TSP was recently introduced as a framework for analyzing signals on higher-order topological spaces, such as simplicial and cell complexes~\cite{barb_2020,barb_Mag_2020,tsp_cell}. Rather than assuming the presence of HOIs \emph{a priori}, TSP provides the minimal topological structure needed to endow edge-centric dynamics with a discrete calculus, enabling the definition of divergence (source–sink organization), gradient (potential-driven flows), and curl (circulatory coordination). In this work, we focus on cell complexes as general models that support these operators while allowing flexible representations of higher-order organization.

A (finite) cell complex (CC) is a combinatorial structure composed of  a collection of basic building blocks, called \emph{cells}. Each cell $c$ has an associated dimension (or order) $k=\mathrm{dim}(c)$, denoted by $c^{k}$, and is referred to as a $k$-cell. For instance, 0-cells, 1-cells and 2-cells correspond to vertices, edges, and polygons, respectively. Unlike simplicial complexes, which restrict 2-cells to triangles (or 2-simplices), CCs allow general polygonal faces (triangles, quadrilaterals, pentagons, etc.), enabling richer and more flexible representations of HOIs. 
How cells of different dimensions are organized and attached to each other is encoded by a boundary (incidence) relation $\prec_b$ that specifies how higher-dimensional cells attach to lower-dimensional ones~\cite{tsp_cell}.  Specifically, if $c^{k-1}\prec_b c^{k}$, then $c^{k-1}$ is a face of $c^{k}$ and lies on its boundary. For example, an edge is bounded by two endpoint faces, while a 2-cell is bounded by an oriented cycle of edges. A CC is $K$-dimensional if all its cells have dimension at most $K$, and all boundary faces of each cell belong to the complex.

A $K$-cell complex is specified by a collection of sets $\{\mathcal{C}^{k}\}_{k=0}^{K}$, where $\mathcal{C}^{k}$ denotes the set of all $k$-cells with cardinality $N_k\triangleq|\mathcal{C}^{k}|$. In order to obtain an algebraic representation of a cell complex, we assign an orientation to every cell. For an incident pair $c_i^{k-1}\prec_b c_j^{k}$, we write $c_i^{k-1}\sim c_j^{k}$ if their orientations are coherent, and $c_i^{k-1}\not\sim c_j^{k}$ otherwise. For edges (1-cells), this means assigning an orientation between endpoints; for 2-cells (polygons), a circulation direction along the boundary, e.g., (counter)clockwise. Given an orientation,  two $k$-cells are said to be lower adjacent if they share a common face of order $k-1$ and upper adjacent if they are both faces of a cell of order $k+1$.
The incidence relations between cells of order $k$ and $k-1$ are encoded by the incidence (or boundary) matrix $\mathbf{B}_k$, whose entries are 1 (or $-1$) if $c_i^{k-1}\prec_b c_j^{k}$ and $c_i^{k-1}\sim c_j^{k}$ (or $\not\sim$), and $0$ otherwise. Therefore, the topology of a $K$-cell complex is fully characterized by the collection of incidence matrices $\{\mathbf{B}_k\}_{k=1}^{K}$. A fundamental property of these matrices is that the boundary of a boundary is empty, i.e., $\mathbf{B}_k\mathbf{B}_{k+1}=\mathbf{0}$ for $k=1,\ldots,K-1$.

\begin{figure}[!t]
\centering
\includegraphics[width=0.7\columnwidth,height=1.75cm]{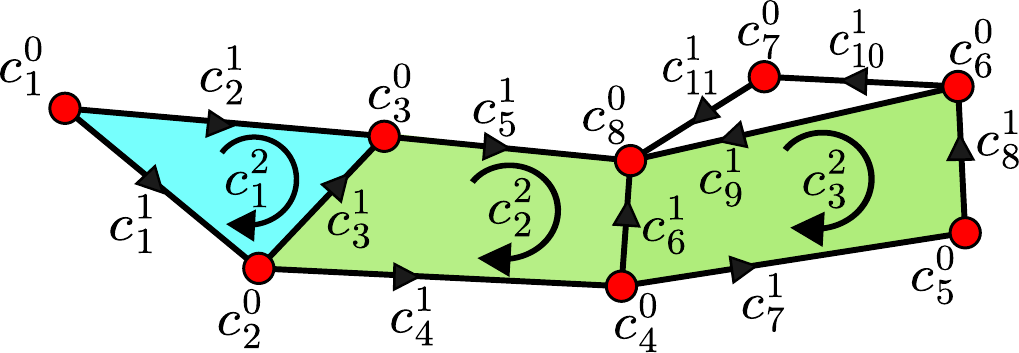}
\caption{An example of a CC of order $2$.}
\label{fig:CMC_1}
\vspace{-0.4cm}
\end{figure}

Throughout this work, we focus on 2-dimensional CCs, denoted by $\mathcal{C}=\{\mathcal{V},\mathcal{E},\mathcal{P}\}$, where $\mathcal{V}=\mathcal{C}^0$, $\mathcal{E}=\mathcal{C}^1$, and $\mathcal{P}=\mathcal{C}^2$ are the sets of vertices, edges, and polygons, with cardinalities $|\mathcal{V}|=N$, $|\mathcal{E}|=E$, and $|\mathcal{P}|=P$, respectively. In this setting, HOIs are represented by polygonal faces (2-cells), excluding higher-dimensional cells (e.g., 3-cells/volumes). The corresponding signed incidence matrices $\mathbf{B}_1\in\mathbb{R}^{N\times E}$ and $\mathbf{B}_2\in\mathbb{R}^{E\times P}$ encode node-edge and edge-face relations, respectively, and provide the algebraic ingredients that later define the Hodge Laplacian, divergence and curl operators.
Fig. \ref{fig:CMC_1} shows an example of a 2-order CC, consisting of $N=8$ nodes (0-cells), $E=11$ edges (1-cells) and $P=3$ polygons (2-cells), namely one triangle, and two squares, along with their corresponding orientations. In our multimodal context of brain networks, vertices correspond to ROIs, edges encode pairwise structural relationships, and faces represent functional HOIs among groups of ROIs. In this work, we restrict the 2-cells to triplet, quadruplet, and quintuplet HOIs.

Based on these operators, the topology of the complex is captured by the combinatorial Hodge Laplacians~\cite{Horak13}.
In particular, the first-order Hodge Laplacian is defined as
\vspace{-0.1cm}
\begin{equation}\label{eq:hogde_laplacian}
  \mathbf{L}_1=\mathbf{B}_1^\top\mathbf{B}_1 + \mathbf{B}_{2}\mathbf{B}_{2}^\top,
\end{equation}
\noindent where $\mathbf{L}_{1,\ell}=\mathbf{B}_1^\top\mathbf{B}_1$ and $\mathbf{L}_{1,u}=\mathbf{B}_{2}\mathbf{B}_{2}^\top$ are the lower and upper Laplacians, encoding adjacency between edges via shared nodes and shared polygons, respectively~\cite{tsp_cell}.\\
\textbf{Spectral signal representation.} 
Within this algebraic framework, a $k$-cell signal is a map $\mathbf{s}^k:\mathcal{C}^k\rightarrow \mathbb{R}^{N_k}$ from the set of cells to the real vectors  $\mathbf{s}^k\in\mathbb{R}^{N_k}$. Accordingly, $\mathbf{s}^0 \,: \,\mathcal{V} \rightarrow \mathbb{R}^N$, $\mathbf{s}^1 \,: \,\mathcal{E} \rightarrow \mathbb{R}^E$, and $\mathbf{s}^2 \,: \,\mathcal{P} \rightarrow \mathbb{R}^P$ denote signals over nodes, edges, and 2-cells, respectively. An orthogonal basis for representing $k$-cell signals is given by the eigenvectors of the $k$-th order Hodge Laplacian $\mathbf{L}_k$, which is a symmetric positive semidefinite matrix. It admits the eigendecomposition $\mathbf{L}_k=\mathbf{U}_k\mathbf{\Lambda}_k\mathbf{U}^{\top}_k$,
where $\mathbf{U}_k$  is the unitary matrix whose columns are the eigenvectors of $\mathbf{L}_k$ and $\mathbf{\Lambda}_k$ is the diagonal matrix whose entries are the associated eigenvalues. The \emph{Cell Fourier Transform} (CFT) of $\mathbf{s}^k$ is defined as $\hat{\mathbf{s}}^k=\mathbf{U}_k^\top\mathbf{s}^k$, allowing $\mathbf{s}^k$ to be expressed in the spectral domain as $\mathbf{s}^k=\mathbf{U}_k\hat{\mathbf{s}}^k$ \cite{tsp_cell}. A signal is said $K$-bandlimited if it can be represented over a set of $K$ eigenvectors. Bandlimitedness in these bases reflects topological smoothness of the signal.\\
\textbf{Hodge decomposition.}
From \eqref{eq:hogde_laplacian} and using the orthogonality condition $\mathbf{B}_1 \mathbf{B}_2=\mathbf{0}$, the space of edge signals decomposes into three orthogonal subspaces, known as the \emph{Hodge decomposition} \cite{Lim}, given by
\vspace{-0.1cm}
\begin{equation}\label{eq:hodge_decomposition}
  \mathbb{R}^{E} \equiv \mathrm{img}(\mathbf{B}_1^\top) \oplus \ker(\mathbf{L}_1) \oplus \mathrm{img}(\mathbf{B}_{2}).
\end{equation}
This implies that any edge signal can be decomposed as
\vspace{-0.1cm}
\begin{equation}\label{eq:Hodge_cell}
\mathbf{s}^1 = \mathbf{B}_1^\top \mathbf{s}^0 + \mathbf{B}_2 \mathbf{s}^2+\mathbf{s}_H,
\end{equation}
\noindent where the three components are mutually orthogonal and admit a physical vector-field interpretation. Specifically,   $\mathbf{s}_{irr}^1=\mathbf{B}_1^\top \mathbf{s}^0$ is
the irrotational edge signal, i.e., the gradient of the node signal across edges; $\mathbf{s}_{s}^1=\mathbf{B}_2 \mathbf{s}^2$ is the solenoidal signal, representing the circulation (curl) along polygons; and $\mathbf{s}_H$ is the harmonic signal lying in both $\text{ker}(\mathbf{B}_1)$ and  $\text{ker}(\mathbf{B}_2^T)$.

Two operators with clear physical interpretations naturally arise in this setting. The divergence operator,
\vspace{-0.1cm}
\begin{equation}\label{eq:div_operator}
\mathrm{div}(\mathbf{s}^1)=\mathbf{B}_1\mathbf{s}^1\in\mathbb{R}^{N},
\end{equation}
yields a node signal that identifies regions acting as sources or sinks of interaction; and the curl operator,
\vspace{-0.1cm}
\begin{equation}\label{eq:curl_operator}
\mathrm{curl}(\mathbf{s}^1)=\mathbf{B}_2^\top\mathbf{s}^1\in\mathbb{R}^{P},
\end{equation}
quantifies the circulation of edge signals along filled polygons. 
By construction, irrotational signals are curl-free, solenoidal signals are divergence-free, and harmonic signals are both, yielding a physically interpretable decomposition that distinguishes source-sink patterns,  cyclic interactions along 2-cells~\cite{bispo}, and topology-constrained global flows. Furthermore, the Hodge decomposition ~\eqref{eq:hodge_decomposition} splits the eigenbasis $\mathbf{U}_1$ into: (i) gradient eigenvectors $\mathbf{U}_{\mathrm{G}}$ spanning $\mathrm{img}(\mathbf{B}_1^\top)$ (associated with nonzero eigenvalues of $\mathbf{L}_{1,\ell}$), (ii) curl eigenvectors $\mathbf{U}_{\mathrm{C}}$ spanning $\mathrm{img}(\mathbf{B}_2)$ (associated with nonzero eigenvalues of $\mathbf{L}_{1,u}$), and (iii) harmonic eigenvectors $\mathbf{U}_{\mathrm{H}}$ spanning $\ker(\mathbf{L}_1)$, whose dimension equals the first Betti number, i.e., the number of independent 1-cycles (unfilled polygons).

\vspace{-0.1cm}
\subsection{Joint  learning of topology and sparse signals}
\label{subsec:joint_learn_alg}

In this section, we present a TSP-based strategy to learn polygonal 2-cells and a sparse signal representation from edge flows (or signals), extending \cite{bispo} from simplicial to cell complexes. Given an underlying graph, the goal is to infer a higher-order topology that enables sparse spectral representations of the solenoidal and harmonic components by jointly minimizing total circulation and data-fitting error.

Let us denote with 
$\hat{\mathbf{s}}^{1}_{s}=\mathbf{U}^{\top}_{\text{C}} \,{{\mathbf{s}}}^1_{s}, \quad \hat{\mathbf{s}}^{1}_{H}=\mathbf{U}^{\top}_{\text{H}} \,{{\mathbf{s}}}^1_{H}$,
the spectral representations of the solenoidal and harmonic signals, respectively.
Given  a graph and observed edge-flow signals $\mathbf{y}^{1}(t) \in \mathbb{R}^{E}$ at time  instants $t=1,\ldots,T$, we build the data matrix $\mathbf{Y}^1=[\mathbf{y}^{1}(1),\ldots, \mathbf{y}^{1}(T)]\in\mathbb{R}^{E\times T}$. From this matrix, our goal is to learn the incidence matrix $\mathbf{B}_2$, or equivalently, the upper Laplacian matrix $\mathbf{L}_{1,u}$. Denoting by $P$ the number of possible 2-cells in the complex, such a matrix can be written in the equivalent form $\mathbf{L}_{1,u}=\sum_{n=1}^{P} q_n \mathbf{b}_n \mathbf{b}_n^\top$, where $\mathbf{b}_n$ are the ($E\times 1$)-dimensional columns of $\mathbf{B}_2$, while $q_n \in\{0,1\}$ are binary  coefficients selecting the filled polygonal cells according to the chosen optimization rule. Then, as in \cite{barb_2020}, we first assess whether the observed signals exhibit solenoidal or harmonic components, as this determines whether it is necessary to move beyond the graph structure and estimate $\mathbf{B}_2$. To do so, we project the observed edge flows onto the subspace orthogonal to the irrotational one
 $\mathbf{y}^{1}_{sH}(t)=(\mathbf{I}-\mathbf{U}_{\text{G}} \mathbf{U}_{\text{G}}^\top) \mathbf{y}^{1}(t)$, $ \forall t$.
We next compute the Frobenius norm of the matrix $\mathbf{Y}^{1}_{sH}=[\mathbf{y}^{1}_{sH}(1), \ldots, \mathbf{y}^{1}_{sH}(T)]$. If this norm exceeds a prescribed threshold, we proceed to estimate $\mathbf{L}_{1,u}$.  

Defining the matrices $\hat{\mathbf{S}}_{s}^{1}=[\hat{\mathbf{s}}^{1}_{s}(1),\ldots, \hat{\mathbf{s}}^{1}_{s}(T)]$, $\hat{\mathbf{S}}_{H}^{1}=[\hat{\mathbf{s}}^{1}_{H}(1),\ldots, \hat{\mathbf{s}}^{1}_{H}(T)]$, we can formulate the non-convex optimization problem
\vspace{-0.2cm}
\begin{equation}
    \label{eq:P_mtv_sign_est1}
\scalebox{0.85}{$
\begin{array}{lll}
\underset{ \substack{\hat{\mathbf{S}}^{1}_{s}, \hat{\mathbf{S}}^{1}_{H}, \mathbf{q} \\ \mathbf{U}_{\text{C}}, \mathbf{U}_{\text{H}},{\boldsymbol{\Lambda}_s}\succeq \mathbf{0}}}{\min} & \!\!\!\!\!\!\displaystyle \sum_{n=1}^{P} q_n \!\parallel \mathbf{Y}^{1\, T}_{sH} \mathbf{b}_n \parallel_F^2\!+\beta \!
\displaystyle  \parallel  \mathbf{Y}^1_{sH}{\scriptstyle -}\mathbf{U}_{\text{C}}\hat{\mathbf{S}}^{1}_{s}{\scriptstyle -}\mathbf{U}_{\text{H}}\hat{\mathbf{S}}^{1}_{H}\parallel_F^2 \\
\quad \text{s.t.} &  \!\!\!\!\!\!a)\, 
\parallel \hat{\mathbf{s}}^{1}_{s}(t) \parallel_1 \leq \alpha_1, \parallel \hat{\mathbf{s}}^{1}_{H}(t) \parallel_1\leq \alpha_2, \; t=1,\ldots,T \\
 \quad \quad & \!\!\!\!\!\!b)\, \displaystyle \sum_{n=1}^{P} q_n \mathbf{b}_n \mathbf{b}_n^{\top} \mathbf{U}_{\text{C}}\!= \!\mathbf{U}_{\text{C}} \boldsymbol{\Lambda}_s, \,  \Lambda_s(i,j)\!=\!0, \forall i \neq j, \\
\quad \quad & \!\!\!\!\!\!c)\, (\displaystyle \sum_{n=1}^{P} q_n \mathbf{b}_n \mathbf{b}_n^{T}+\mathbf{B}_1^\top \mathbf{B}_1) \mathbf{U}_{\text{H}}= \mathbf{0}, \\
\quad \quad & \!\!\!\!\!\!d)\, \parallel \mathbf{q} \parallel_0 \leq q^{\star}, \quad \mathbf{q} \in \{0,1\}^{P}
\end{array}
$}
\end{equation}
where  the first term in the objective function represents the energy of the signals circulation along all $2$-order cells, while the second term accounts for the data-fitting weighted  by the coefficient $\beta>0$. The $l_1$-norm constraints in  $a)$ promote spectral sparsity of the signals through the positive coefficients $\alpha_1,\alpha_2$; 
$b)$ and $c)$ enforce that the matrices $\mathbf{U}_{\text{C}}$ and $\mathbf{U}_{\text{H}}$ correspond to the eigenvectors associated with the non-zeros eigenvalues  of $\mathbf{L}_{1,u}$, and to the eigenvectors spanning the kernel of $\mathbf{L}_1$, respectively. Finally, the constraint $d)$ ensures that the number of cells of $2$-order does not exceed the maximum number $q^{\star}$. Since \eqref{eq:P_mtv_sign_est1} is a non-convex problem due to the constraints $b)$, $c)$ and $d)$, we propose an iterative sub-optimal algorithm  which alternates between the solutions of two convex optimization problems for each possible number $q^{\star}=1,\ldots,P$ of polygonal cells.
Specifically,  we iterate for $i=1, \ldots, P$ between the following two steps:
\begin{itemize}
    \item[S.1] 
    Solve \eqref{eq:P_mtv_sign_est1} with respect to $\mathbf{q}$, setting $\beta=0$, $q^{\star}=i$ and considering only constraints $b)$, $c)$ and $d)$.
    This problem admits a closed-form  solution that can be obtained by computing $w_n= ||\mathbf{Y}^{1\, \top}_{sH} \mathbf{b}_n||_F^2$ and sorting these coefficients  in increasing order. Then, selecing the indices $\mathcal{P}^{\star}=\{i_1,i_2,\ldots,i_{q^{\star}}\}$ of the $q^{\star}$ lowest coefficients $w_n$, we get $\tilde{\mathbf{L}}_{1,u}^{i}=\sum_{n\in \mathcal{P}^{\star}} \mathbf{b}_n \mathbf{b}_n^\top$;
    \item[S.2] Given $\tilde{\mathbf{L}}_{1,u}^{i}$, find the eigenvector matrices $\mathbf{U}_{\text{C}}^{i}$  and, from the  estimated Hodge Laplacian $\tilde{\mathbf{L}}_{1}^{i}=\mathbf{L}_{1,\ell}+\tilde{\mathbf{L}}_{1,u}^{i}$, derive the eigenvector matrix $\mathbf{U}_{\text{H}}^{i}$  spanning its kernel. Then, the convex optimization problem \eqref{eq:P_mtv_sign_est} is solved.
    \vspace{-0.05cm}
    \begin{equation}  
\label{eq:P_mtv_sign_est}
\hspace{-0.1cm}
\scalebox{0.85}{$
\begin{array}{lll}
\underset{ \hat{\mathbf{S}}^{1}_{s}, \hat{\mathbf{S}}^{1}_{H}}{\min} & \displaystyle \parallel \mathbf{Y}^1_{sH}{\scriptstyle -}\mathbf{U}_{\text{C}}^{i}\hat{\mathbf{S}}^{1}_{s}{\scriptstyle -}\mathbf{U}_{\text{H}}^{i}\hat{\mathbf{S}}^{1}_{\text{H}}\parallel_F^2 \quad \\
\quad \text{s.t.} &  a) 
\parallel \hat{\mathbf{s}}^{1}_{s}(t) \parallel_1 \leq \alpha_1, \;  \parallel \hat{\mathbf{s}}^{1}_{H}(t) \parallel_1\leq \alpha_2, \; t=1,\ldots,T. \medskip\\
\end{array}
$}
\end{equation}

\end{itemize}
\vspace{-0.15cm}
At each iteration $i$, we find the optimal data fit error $\text{g}(i)=||\mathbf{Y}^1_{sH}-\mathbf{U}_{\text{C}}^{i}\hat{\mathbf{S}}^{1}_{s}-\mathbf{U}_{\text{H}}^{i}\hat{\mathbf{S}}^{1}_{\text{H}}||_F^2$ by alternating between steps $\text{S.1}$ and $\text{S.2}$. The  2-cells are then identified as   $\bar{q}^{\star}= \arg\min_{i \in \{1,\ldots, P\}} \text{g}(i)$ and the learned Laplacian is obtained as $\tilde{\mathbf{L}}_{1}=\mathbf{L}_{1,\ell}+\tilde{\mathbf{L}}_{1,u}^{\bar{q}^{\star}}$, using the corresponding index set $\mathcal{P}^{\star}$.

\section{Learning brain topology from data}
\label{sec:cell_learning}

In this study, we adopt a structural-functional learning framework to infer subject-specific brain 2-cell complexes by integrating dMRI-derived structural connectivity with rs-fMRI-derived functional dynamics in 100 unrelated healthy adults from the \href{http://humanconnectome.org/}{HCP} (ages 21-35; 46 males/54 females)~\cite{hcp_dataset}. Brains were parcellated with the Automated Anatomical Labeling (AAL) atlas~\cite{TzourioMazoyer2002AAL} into $N=90$ ROIs, each assigned to one of seven canonical functional subnetworks (FS) and its corresponding colormap~\cite{parcellation}: frontoparietal (FP, orange), default mode network (DMN, red), subcortical (SC, gray), visual (VIS, purple), somatomotor (SM, steelblue), ventral attention (VA, violet), and limbic (LIM, khaki).

For each subject, we infer the brain 2-CC in three steps: (i) identify statistically significant structural edges, (ii) construct functional edge signals, and (iii) learn HOCs and sparse signal representations under structural-functional constraints.

\noindent\textbf{Inferring significant structural edges.}
For each subject, dMRI data yielded a weighted structural adjacency matrix whose edge weights reflect white-matter fiber densities between pairs of ROIs. Statistical significance was assessed using 5000 degree-preserving Rubinov-Sporns rewired null networks \cite{rubinov2011}; each empirical edge weight was compared to its null distribution via a Student’s $t$-test, retaining connections with $p<.05$ \cite{null_models}. This yields a sparse binary structural adjacency matrix ($\approx$5\% density), from which we build the node-edge incidence matrix $\mathbf{B}_1^{(S)}\in\mathbb{R}^{N\times E^{(S)}}$ with entries $\{0,\pm1\}$, for each subject $S$. Since edge orientations are arbitrary, we use a lexicographic convention: an edge $e_{i,j}$,  connecting ROIs $v_i$ and $v_j$, is oriented as $v_i \rightarrow v_j$ if $i < j$. HOC orientations are then fixed by their subject-specific boundary matrices, with no degrees of freedom beyond a global sign convention.

\noindent\textbf{Defining functional edge signals.}
To enable TSP-based learning and analysis, functional signals are defined on the edges of the structural backbone. Let 
\(
\mathbf{s}^0(t) = [s^0_{v_1}(t), \ldots, s^0_{v_N}(t)]^\top
\)
denote the node signal of z-scored rs-fMRI activities across the $N$ ROIs $v_i$ at time $t$.
Time-resolved edge signals are commonly derived from amplitude co-fluctuations \cite{Faskowitz2022, Zamani2022} or phase-based coupling measures \cite{Cabral2017,leida}. 
Here, we use instantaneous phase-coherence, a bounded time-resolved measure in [-1,1] (near ±1: strong phase (anti-)synchronization; near 0: weak coupling), suitable for flow-based TSP analysis and provides a stable, comparable scale across subjects. 
Specifically, for an edge $e_{i,j}$, the edge signal at time $t$ is defined as
\vspace{-0.1cm}
\begin{equation}\label{eq:phase_coherence}
s^1_{e_{i,j}}(t) = \cos\!\left( \phi(s^0_{v_i}(t)) - \phi(s^0_{v_j}(t)) \right),
\end{equation}
where $\phi(\cdot)$ denotes the instantaneous phase extracted via the Hilbert transform~\cite{Cabral2017, leida}.
These edge signals allow us to define a time-varying vector field over the brain network.

\noindent\textbf{Joint learning of higher-order cells.}
HOCs are inferred using the learning method in Sec.~\ref{subsec:joint_learn_alg}, which integrates the sparse structural scaffold $\mathbf{B}_1^{(S)}$ with functional edge signals $\mathbf{y}^1(t)$ from \eqref{eq:phase_coherence} to identify the minimal set of 2-cells (triangles, quadrilaterals and pentagons) that best explains the data. For each subject $S\in\{1,\ldots,100\}$, the procedure estimates the edge-polygon incidence matrix $\tilde{\mathbf{B}}_2^{(S)}$ and the associated first-order Hodge Laplacian $\tilde{\mathbf{L}}_{1}^{(S)}=\mathbf{L}_{1,\ell}^{(S)}+\tilde{\mathbf{L}}_{1,u}^{(S)}$, yielding a subject-specific 2-cell complex $\mathcal{C}^{(S)}=\{\mathcal{V},\mathcal{E}^{(S)},\mathcal{P}^{(S)}\}$. The vertex set $\mathcal{V}$ is fixed across subjects ($N=90$ ROIs), whereas $\mathcal{E}^{(S)}$ and $\mathcal{P}^{(S)}$ are subject-dependent, capturing individual variability in anatomical-functional organization. The 2-cell set consists of $\mathcal{P}^{(S)}=\{\mathcal{P}_3^{(S)},\mathcal{P}_4^{(S)},\mathcal{P}_5^{(S)}\}$, with $|\mathcal{P}_{n=3,4,5}^{(S)}|=P_{n=3,4,5}^{(S)}$, enabling individual-level topological analysis of brain dynamics while explicitly capturing higher-order structure.

We emphasize that all HOCs are inferred from pairwise structural-functional interactions, without assuming intrinsically higher-order statistical dependencies. They provide a topological organization of edge dynamics, rather than independent higher-order measurements, and form the minimal higher-order scaffold required for sparse spectral representations of rotational signals not expressible in graphs alone. 
Furthermore, because edge orientations in each subject-specific topology are assigned arbitrarily for mathematical consistency, with 2-cell orientations fixed by the boundary matrices, divergence sign (positive: source; negative: sink) and circulation direction (positive: counterclockwise; negative: clockwise) are convention-dependent and thus have no intrinsic physical meaning in fMRI data. Accordingly, sign-sensitive interpretations must be understood relative to the adopted convention. Therefore, all results in this work rely on orientation-invariant quantities—such as magnitudes, energy distributions across gradient and rotational subspaces, and temporal statistics—ensuring that our findings are independent of the specific orientation choice.
\vspace{-0.1cm}
\section{Results and discussion}\label{sec:results}

This section describes the learned subject-specific brain 2-CCs and their dynamics, from hierarchical structural-functional organization and dominant gradient-curl flow regimes to significant topological brain-behavior associations.
\vspace{-0.6cm}
\subsection{Structural-functional organization of brain topologies}
\label{subsec:mesoscale}

\begin{figure}[!t]
\centering

\subfigure[]
{
\includegraphics[width=0.85\columnwidth,height=3.0cm]{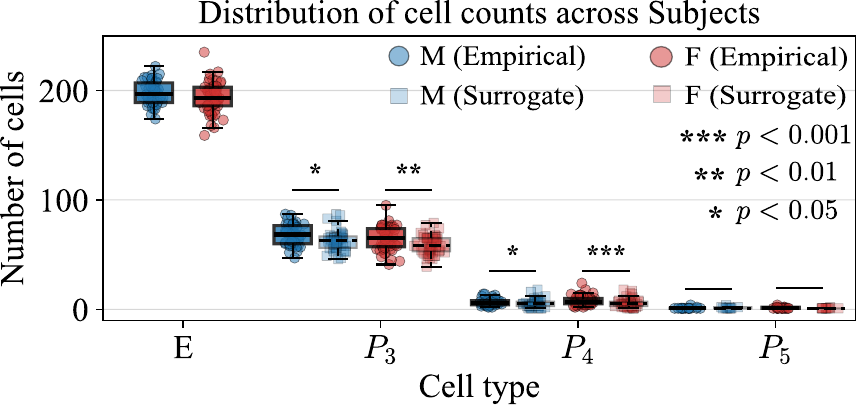}
\label{subfig:cell_counts}
}
\vspace{-0.1cm}
\\
\subfigure[]
{
\includegraphics[width=0.85\columnwidth,height=2.9cm]{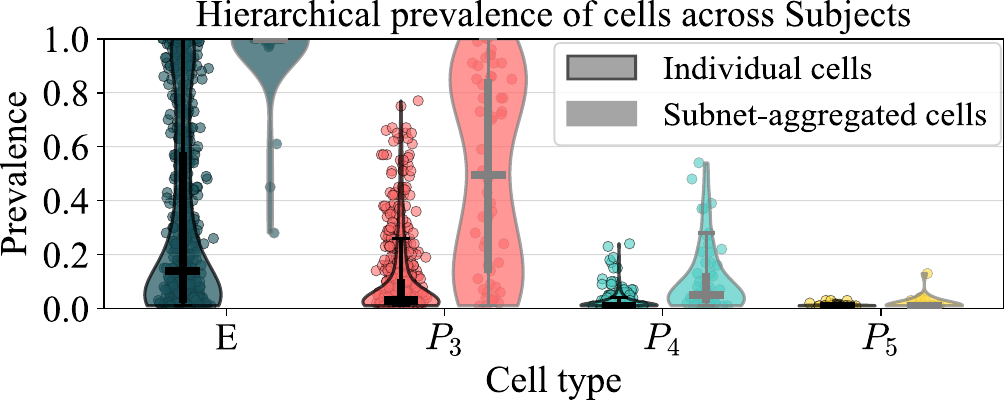}
\label{subfig:prevalence}
}
\vspace{-0.1cm}
\\
\subfigure[]
{
\includegraphics[width=0.47\columnwidth,height=3.2cm]{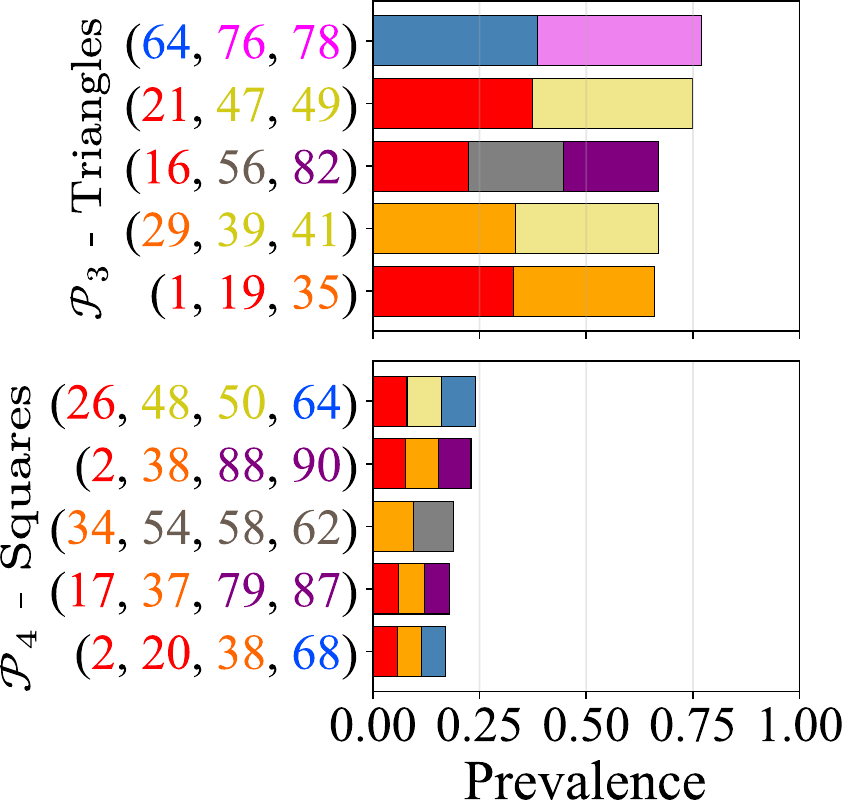}
\label{subfig:top_cell_prev}
}
\hspace{-0.25cm}
\subfigure[]
{
\includegraphics[width=0.47\columnwidth,height=3.2cm]{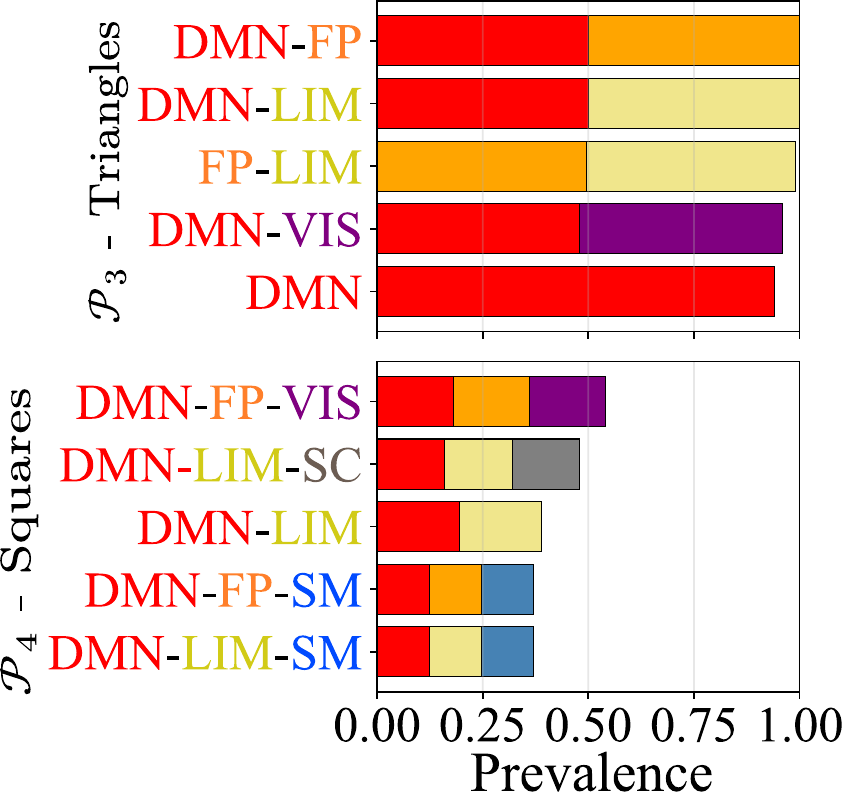}
\label{subfig:top_subnet_prev}
}
\caption{Hierarchical organization and inter-subject consistency of learned brain 2-CCs. \subref{subfig:cell_counts} Subject-specific counts of edges ($E$) and HOCs (triangles $P_3$, quadrilaterals $P_4$, and pentagons $P_5$). \subref{subfig:prevalence} Cell- and subnet-level prevalence distributions across subjects; 3D brain visualizations with ROIs at atlas Euclidean centroids are provided in the Supplementary Material. \subref{subfig:top_cell_prev} Five most prevalent node-level triangles and squares. 
\subref{subfig:top_subnet_prev} Five most prevalent subnet combinations across subjects.}
\label{fig:subject_topologies}
\vspace{-0.3cm}
\end{figure}

We quantitatively characterize subject-specific brain topologies in terms of the number and composition of edges and HOCs. 
Fig.~\ref{subfig:cell_counts} summarizes edge and HOC count distributions across subjects (one point per subject), with color indicating sex (blue: male; red: female). On average, subjects had $196\pm13$ edges ($[159,235]$), $67\pm11$ triangles ($[41,95]$), $8\pm4$ squares ($[0,24]$), and $2\pm1$ pentagons ($[0,4]$). Mann-Whitney tests revealed no significant sex differences for any HOC type ($p>0.05$), indicating no systematic sex dependence of global cell distributions in this healthy young-adult cohort.
Furthermore, to test whether the inferred HOIs reflect structured functional dynamics rather than random temporal fluctuations, we repeated the HOC learning by keeping the same subject-specific structural graph and using phase-randomized edge signals, preserving the signal amplitude spectrum and marginal temporal structure while disrupting cross-edge phase alignment~\cite{phase_randomization}. Since the structural graphs were already obtained by a statistically validated procedure, this analysis was restricted to the inferred HOCs.
The surrogate HOC distributions (transparent square markers in Fig.~\ref{subfig:cell_counts}) differ significantly from the empirical ones for triangles and quadrilaterals ($p<.05$), indicating that these HOIs are not explained by chance pairwise temporal fluctuations, but arise from structured coordination across edges that is disrupted by the phase-randomized null model. In contrast, pentagon counts do not differ significantly from surrogates, likely because higher-cardinality 2-cells are rarer and harder to infer reliably under sparse structural backbones.

To assess inter-subject consistency, we computed \emph{cell prevalence}, defined as the fraction of subjects in which an identical cell (exact node composition) appears, capturing reproducibility at the finest (node-matched) scale. Fig.~\ref{subfig:prevalence} (solid colors) shows a clear hierarchy, where edges exhibit the highest prevalence, whereas HOCs become progressively less prevalent with increasing HOC's cardinality. Triangles ($P_3$) already display marked variability, while quadrilaterals ($P_4$) and pentagons ($P_5$) become increasingly sparse, highlighting the individualized, dynamic nature of functional HOIs beyond fixed structural scaffolding.

To move beyond strict node-level matching and better capture shared mesoscale structure, we analyzed \emph{subnet-combination prevalence}, defined as the fraction of subjects with at least one HOC for a given combination of FSs among its constituent ROIs. The transparent plots in Fig.~\ref{subfig:prevalence} show that this mesoscale aggregation markedly increases prevalence across all cell types, revealing a more stable and recurrent organization across subjects up to quadrilaterals.

Despite cell-level variability, the most prevalent HOCs reveal structured organization. Fig.~\ref{subfig:top_cell_prev} shows the five most prevalent triangular and quadrilateral cells across subjects, with bar length indicating prevalence and colors denoting the involved FSs. Dominant triangles primarily link transmodal ROIs (DMN/FP), often involving LIM, whereas prevalent quadrilaterals show more heterogeneous compositions, combining transmodal, unimodal (SM/VIS), and LIM regions.
The predominance of transmodal-limbic motifs suggests that these systems recurrently participate in HOIs consistently expressed across individuals. The prominence of transmodal-limbic motifs aligns with the integrative role of transmodal cortex, particularly the DMN, along cortical hierarchies and macroscale functional gradients \cite{van2013network}, while the broader mixing in quadrilaterals aligns with hierarchical models in which unimodal systems anchor lower gradients and transmodal systems support cross-modal integration \cite{Margulies2016Gradient}.

The most prevalent subnet combinations (Fig.~\ref{subfig:top_subnet_prev}) corroborate the cell-level results. For triangular HOCs, the DMN appears in nearly all subjects, either within-network or coupled with FP and LIM, consistent with their roles in resting-state processes—self-referential cognition (DMN), cognitive flexibility (FP), and emotion integration (LIM)~\cite{stephen, parcellation}. Quadrilateral HOIs show stronger transmodal-unimodal mixing, suggesting that higher-cardinality cells support more distributed cross-hierarchical integration beyond the core resting-state backbone. Recurrent DMN-VIS HOCs further indicate structured coupling between internally oriented cognition and perceptual systems, consistent with evidence that VIS also supports internally generated processes such as mental imagery, episodic memory, and spontaneous thought at rest \cite{AndrewsHanna2014DMN, Zhang2018Intrinsic}.

To quantify inter-subject consistency across scales, we computed Jaccard similarity between subject-specific topologies at both the cell and subnet levels. Here, the Jaccard index is the intersection-over-union of two subjects’ cell sets (or subnet-combinations), with values near 1 (or 0) indicating high (or low) overlap. Group-level scores were obtained by averaging the upper-triangular entries of the pairwise Jaccard matrices.

At the microscopic (cell) level, pairwise structural edges show moderate inter-subject overlap (mean Jaccard-index $\mathtt{J}_\mathcal{E}=0.542$), coherent with relatively conserved nature of anatomical connectivity across healthy individuals. In contrast, HOIs are far less shared: triangles yield $\mathtt{J}_{\mathcal{P}_3}=0.175$, whereas quadrilaterals and pentagons are near-zero ($\mathtt{J}_{\mathcal{P}_4}=0.030$, $\mathtt{J}_{\mathcal{P}_5}=0.013$). The decreasing size of the Jaccard similarity matrices (100 subjects for $\mathcal{P}_3$, 99 for $\mathcal{P}_4$, and 34 for $\mathcal{P}_5$) mirrors the reduced prevalence of HOCs (Fig.~\ref{subfig:prevalence}) and indicates that node-based HOCs are highly individualized.

After aggregating cells by FS composition, inter-subject similarity increases markedly across all orders. At the subnet level, edges become nearly identical across subjects ($\mathtt{J}_{\mathcal{E}}=0.972$), and HOCs show substantially higher overlap ($\mathtt{J}_{\mathcal{P}_3}=0.607$, $\mathtt{J}_{\mathcal{P}_4}=0.145$, $\mathtt{J}_{\mathcal{P}_5}=0.123$). Overall, higher-order topology is strongly scale-dependent, specifically, node-based motifs are individualized, whereas mesoscale subnet organization is more conserved and reproducible.

\vspace{-0.1cm}
\subsection{Dominant gradient and rotational flow organization}\label{subsec:spectral_components}

Our goal is to identify, through the Hodge-spectral decomposition, significant topological flow regimes that persist throughout the resting-state acquisition and are supported by the learned anatomical-functional scaffold.

For each subject $S$, we consider the first-order Hodge Laplacian $\mathbf{L}_1^{(S)}$ associated with the learned 2-cell complex. Rather than analyzing the full eigenbasis $\mathbf{U}_1^{(S)}$, we exploit the Hodge decomposition in \eqref{eq:Hodge_cell}, separating the edge signals into orthogonal gradient and rotational flow subspaces spanned by $\mathbf{U}_{\mathrm{G}}^{(S)}$ and $\mathbf{U}_{\mathrm{C}}^{(S)}$, respectively.
Given the time-varying edge signals $\mathbf{s}^{1\,(S)}(t)\in\mathbb{R}^{E^{(S)}}$ from rs-fMRI phase-coherence \eqref{eq:phase_coherence}, we characterize gradient and rotational dynamics using sparse representations. For the {gradient} component, we project $\mathbf{s}^{1\,(S)}(t)$ onto the gradient Hodge subspace through the CFT coefficients $\hat{\mathbf{s}}_{\mathrm{G}}^{1\,(S)}(t)=\mathbf{U}_{\mathrm{G}}^{(S)\top}\mathbf{s}^{1\,(S)}(t)$. For each gradient spectral index, we compute its time-averaged power spectral density (PSD) and assess relevance via a subject-specific surrogate baseline: we generate 5000 phase-randomized edge signals realizations and build a null distribution of PSD values for each spectral index. We retain only the high-energy spectral indices whose empirical PSD is significantly above the null baseline ($p<.05$), yielding the index set $\mathcal{I}_{\mathrm{G}}^{(S)}$. The filtered gradient flow is then reconstructed as \( \mathbf{s}_{\mathrm{G}}^{(S)}(t)= \mathbf{U}_{\mathrm{G},\mathcal{I}_{\mathrm{G}}}^{(S)} \mathbf{U}_{\mathrm{G},\mathcal{I}_{\mathrm{G}}}^{(S)\top} \mathbf{s}^{1\,(S)}(t). \)
For the {rotational} part, we use the sparse solenoidal representation returned by the joint learning method (Sec.~\ref{subsec:joint_learn_alg}) to reconstruct $\mathbf{s}_{\mathrm{C}}^{(S)}(t)$. 
We summarize persistent structure by averaging the temporal mean magnitude of $\mathbf{s}_{\mathrm{G}}^{(S)}(t)$ and $\mathbf{s}_{\mathrm{C}}^{(S)}(t)$ across subjects, and retain the top 1\% highest-magnitude edge flows.

\begin{figure}[!t]
\centering

\subfigure[]
{
\includegraphics[width=0.8\columnwidth, height=2.9cm]{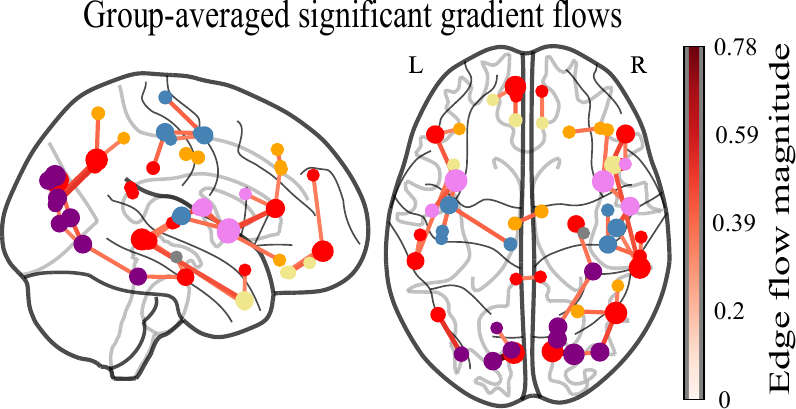}
\label{subfig:mean_significant_gradient_flow}
}
\vspace{-0.1cm}
\\
\subfigure[]
{
\includegraphics[width=0.8\columnwidth, height=2.9cm]{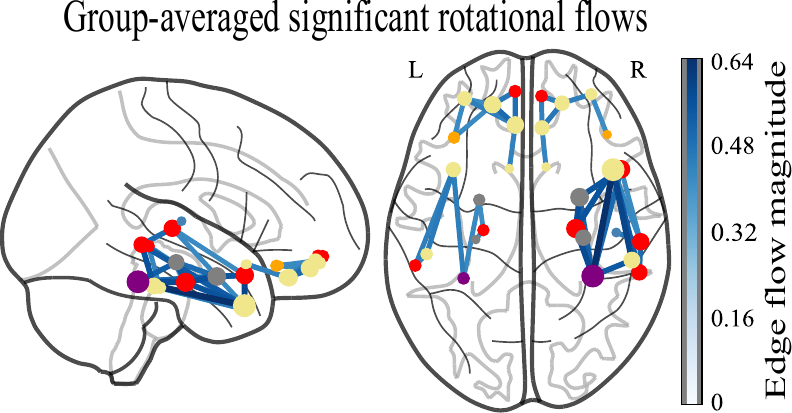}
\label{subfig:mean_significant_curl_flow}
}
\vspace{-0.1cm}
\\
\subfigure[]
{
\includegraphics[width=0.8\columnwidth, height=4.5cm]{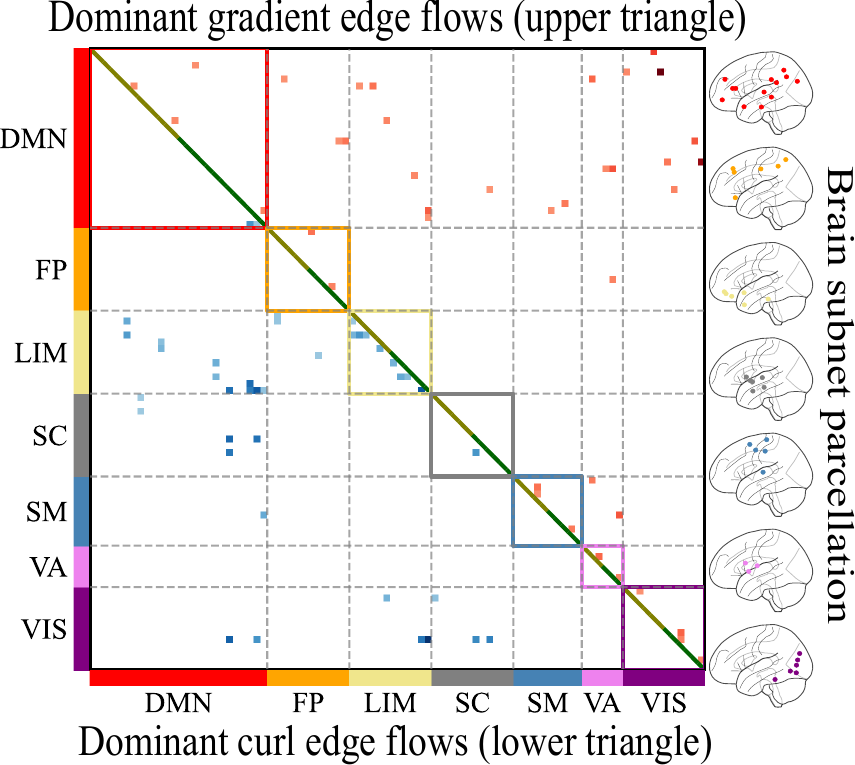}
\label{subfig:heatmap_dominant_gradient_curl_flows}
}

\caption{Dominant gradient and rotational flows: \subref{subfig:mean_significant_gradient_flow}-\subref{subfig:mean_significant_curl_flow} show the group-averaged dominant gradient and rotational flows, respectively. Node color denotes FS membership, node size scales with degree, and edge color intensity represents mean flow magnitude across subjects (red: gradient; blue: rotational); 3D visualizations are available in the Supplementary Materials.
\subref{subfig:heatmap_dominant_gradient_curl_flows} summarizes both components in a matrix view (gradient/rotational in above/below diagonal), with hemispheric boundaries marked on the diagonal (left/right: olive/dark green color).
}
\label{fig:dominant_flows}
\vspace{-0.4cm}
\end{figure}

The dominant gradient flows (Fig.~\ref{subfig:mean_significant_gradient_flow}) reveal a spatially distributed backbone centered on the DMN. Significant gradient edges link DMN to VIS, LIM, FP, VA, SM, and SC systems, forming a large-scale interaction scaffold extending along a posterior-anterior axis from occipital visual areas to prefrontal association cortex. Within the TSP framework, gradient components correspond to source-sink imbalances aligned with node potentials. Their spatial continuity and cross-system distribution are consistent with macroscale cortical hierarchies, in which unimodal sensory regions anchor posterior regions and transmodal systems support integrative coordination~\cite{Margulies2016Gradient, huntenburg2018}. The prominent involvement of the DMN further supports its role as a transmodal hub at the apex of cortical hierarchies, coordinating distributed resting-state dynamics and internally oriented integration~\cite{menon2023dmn}.
In contrast, dominant rotational flows (Fig.~\ref{subfig:mean_significant_curl_flow}) are localized and circuit-specific, with the strongest significant edges clustering bilaterally within limbic regions via intra-limbic interactions and limbic couplings with DMN, SC, and VIS. Within the Hodge framework \eqref{eq:Hodge_cell}, these components capture curl-supported circulation on 2-cells and thus cannot be reduced to gradients of node potentials, indicating localized cyclic dynamics embedded in HOIs. Relative to the spatially distributed gradient backbone, these patterns suggest recurrent limbic-centered coordination linked to internally oriented cognitive and affective functions, consistent with the role of LIM-DMN-SC circuits in emotional regulation, memory integration, prospection, and internally generated thought \cite{AndrewsHanna2014DMN}. 
The recurrent involvement of VIS subnet is consistent with resting-state roles of visual cortex in mental imagery and episodic memory \cite{AndrewsHanna2014DMN,Zhang2018Intrinsic}.
Some strong rotational motifs link orbitofrontal limbic regions with frontal control regions, suggesting that cyclic dynamics extend beyond a purely limbic module into a broader transmodal motif. Consistent with evidence that orbitofrontal/anterior temporal regions belong to an extended default-network architecture and couple to frontoparietal control systems, these curl-supported motifs may reflect recurrent coordination between internally oriented cognition, affective valuation, and executive control \cite{lim_dmn}.

Fig.~\ref{subfig:heatmap_dominant_gradient_curl_flows} summarizes the dominant significant flows, showing gradient-dominated edges above the diagonal and rotational-dominated edges below it. The heatmap highlights a clear spatial dissociation: gradient edges are broadly distributed across functional systems and hemispheres, prominently involving DMN regions and extensive inter-network coupling, consistent with a large-scale integrative backbone. In contrast, rotational edges are more localized within LIM-SC and LIM-DMN blocks with additional VIS couplings, indicating curl-supported dynamics embedded in specific mesoscale circuits rather than global redistribution.

\vspace{-0.1cm}
\subsection{Source/sink organization and behavioral variability}
\label{subsec:divergence_results}

\begin{figure}[!t]
\centering
\subfigure[]
{
\includegraphics[width=0.8\columnwidth,height=2.5cm]{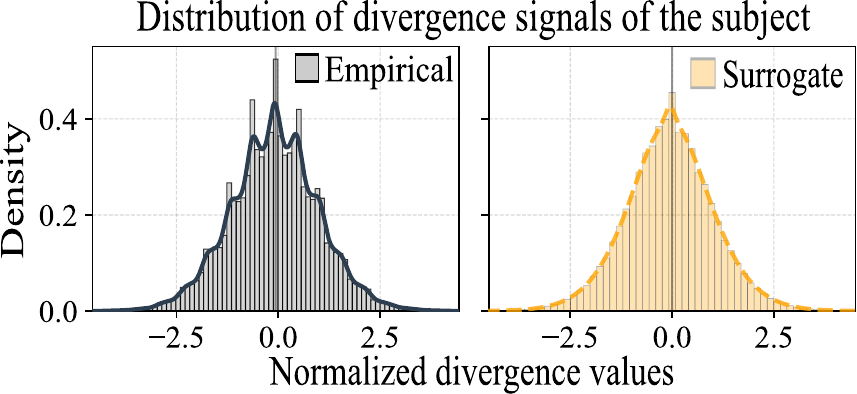}
\label{subfig:div_distribution}
}
\vspace{-0.1cm}
\\
\subfigure[]
{
\includegraphics[width=0.9\columnwidth,height=3.0cm]{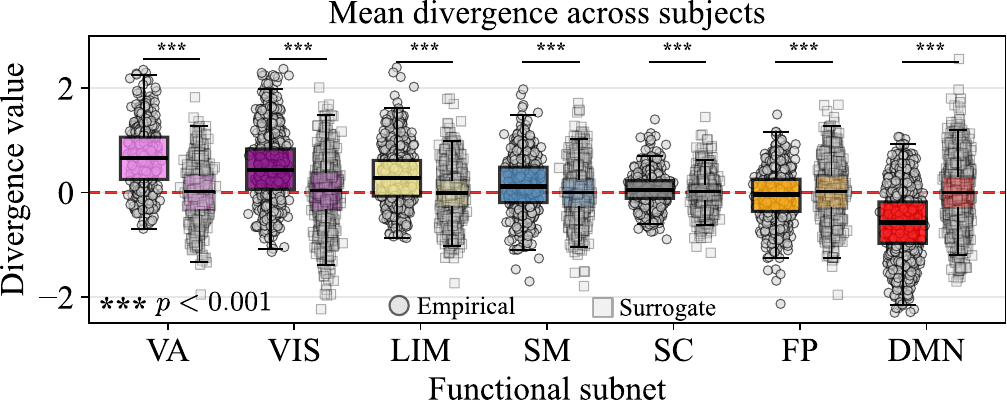}
\label{subfig:divergence_patterns}
}
\vspace{-0.1cm}
\\
\subfigure[]
{
\includegraphics[width=0.46\columnwidth,height=2.5cm]{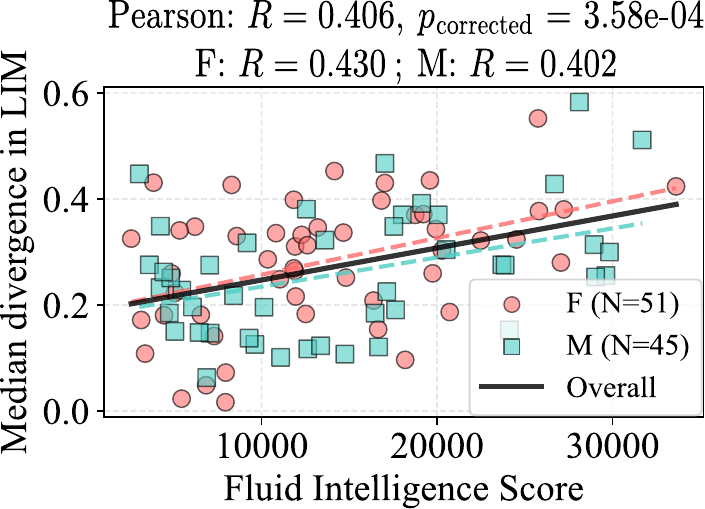}
\label{subfig:correlation_plot_div_lim_pmat24}
}
\subfigure[]
{
\includegraphics[width=0.46\columnwidth,height=2.5cm]{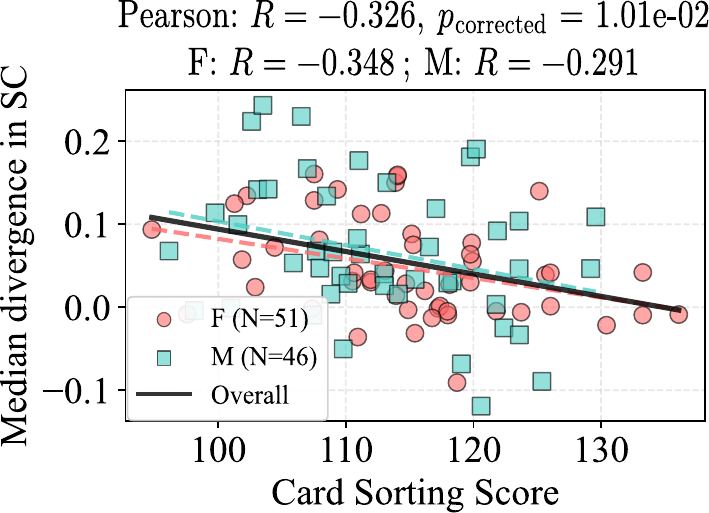}
\label{subfig:correlation_plot_div_sc_cardsort}
}
\caption{{Mesoscale source-sink organization and behavioral associations:}
\subref{subfig:div_distribution} Empirical and surrogate z-scored divergence distributions across subjects;
\subref{subfig:divergence_patterns} time-averaged divergence by FS, shown as boxplots (median, IQR; whiskers $1.5\times$IQR) with overlaid data points; brain-behavior association between \subref{subfig:correlation_plot_div_lim_pmat24} LIM divergence and fluid intelligence; and \subref{subfig:correlation_plot_div_sc_cardsort} SC divergence and executive flexibility (Card Sort). Outliers were removed from association analyses using the $1.5\times$IQR rule.
}
\vspace{-0.4cm}
\label{fig:clinical_div}
\end{figure}

To characterize persistent source-sink organization, we computed for each subject $S$ the time-resolved divergence of the edge-flow signals as $\mathtt{div}(\mathbf{s}^{1\,(S)}(t))=\mathbf{B}_1^{(S)}\mathbf{s}^{1\,(S)}(t)\in\mathbb{R}^{N\times T}$, which quantifies whether each node exhibits a net outward (source-like) or inward (sink-like) redistribution bias over the learned structural scaffold. Fig.~\ref{subfig:div_distribution} shows the z-scored divergence distribution across subjects, departing from the approximately Gaussian surrogate profile and indicating non-random temporal structure. 
These deviations are further amplified in the time-averaged divergence 
(Fig.~\ref{subfig:divergence_patterns}), where empirical FS-grouped 
distributions differ significantly from surrogates across all FSs 
($p<10^{-3}$), evidencing statistically significant mesoscale 
source-sink organization.

A clear subsystem-dependent divergence gradient emerges (Fig.~\ref{subfig:divergence_patterns}). The VA, VIS, and LIM networks show consistently positive median divergence, whereas the DMN exhibits a pronounced negative shift, mirroring the canonical hierarchical gradient of cortical organization spanning unimodal sensory regions to transmodal association cortex \cite{Margulies2016Gradient, huntenburg2018}. VIS positivity may reflect structured intrinsic resting activity \cite{Zhang2018Intrinsic}. VA positivity is consistent with its role in salience detection and attentional reorienting, supporting mesoscale resource reallocation without implying directed signaling \cite{Corbetta2002}. LIM positivity, in turn, may reflect its integrative contribution linking motivational and valuation processes to distributed cortical coordination \cite{Pessoa2017,lim_dmn_hubs}. On the other hand, the DMN’s negative divergence indicates an inward integrative configuration consistent with its role as a transmodal hub coordinating cross-system inputs during internally oriented cognition \cite{menon2023dmn}.
Interestingly, SC divergence is tightly centered near zero with low inter-subject variability, congruent with subcortical (thalamic/striatal) circuits primarily mediating gating, relay, and switching between cortical states rather than persistently driving large-scale propagation \cite{Sherman2017, cruz}. This near-zero profile suggests a dynamically balanced mediation role within cortico-striato-thalamic circuits that support state transitions and executive control.

We next explored whether subject-specific divergence profiles relate to inter-individual cognitive variability by correlating subsystem-specific divergence metrics with some HCP behavioral measures~\cite{Barch2013}, including fluid intelligence, executive function, working memory, emotional processing, motor dexterity, and language/vocabulary performance.
Throughout this work, we focus on statistically significant topological brain–behavior associations with $|R|>0.3$ and Bonferroni-corrected $p_{\textbf{corrected}}<.05$ (across nine behavioral variables), consistent with recent HCP brain–behavior findings \cite{santoro2024higher,Wang2025}.

{Here, we highlight two behavioral measures, namely fluid intelligence, indexing abstract reasoning and relational problem solving, and Card Sort performance, indexing executive flexibility through adaptive rule switching and updating \cite{Barch2013}. Fig.~\ref{subfig:correlation_plot_div_lim_pmat24} shows a significant positive association between median LIM divergence and fluid intelligence (overall $R=0.408$), where each marker represents one subject (circles: females; squares: males), with comparable effects in females and males (both $R>0.40$), indicating robustness across sexes. 
Although limbic circuits are not core FP components, prior work suggests that limbic–cortical interactions support motivational salience, valuation, and sustained engagement relevant to higher-order cognition \cite{Pessoa2017}. Thus, the observed association may reflect co-occurring organized limbic redistribution and more efficient coordination of motivational–valuation processes with large-scale associative networks, potentially supporting abstract reasoning without implying causality. 
In contrast, SC divergence is negatively associated with Card Sort performance ($R=-0.314$; Fig.~\ref{subfig:correlation_plot_div_sc_cardsort}), suggesting that greater subcortical flow imbalance may reflect reduced switching efficiency underlying executive flexibility, aligned with the role of thalamic/striatal circuits in gating, rule updating, and dynamic cortical coordination \cite{Sherman2017,cruz}.}

\vspace{-0.1cm}
\subsection{Circulation organization and behavioral variability}
\label{subsec:curl_results}

Extending the divergence analysis, we next examined the rotational component of functional organization, quantified by the curl operator in \eqref{eq:curl_operator}, which isolates closed-loop coordination patterns over HOIs that cannot be reduced to pairwise interactions. For each subject $S$, time-resolved circulation values were computed over all learned 2-cells as
\(
\mathtt{curl}(\mathbf{s}^{1\,(S)}(t))=\mathbf{B}_2^{\top (S)} \mathbf{s}^{1\,(S)}(t)\in\mathbb{R}^{P^{(S)}\times T},
\)
thereby capturing cyclic coordination motifs supported by the higher-order scaffold. 

\begin{figure}[!t]
\centering
\subfigure[]
{
\includegraphics[width=0.8\columnwidth,height=2.5cm]{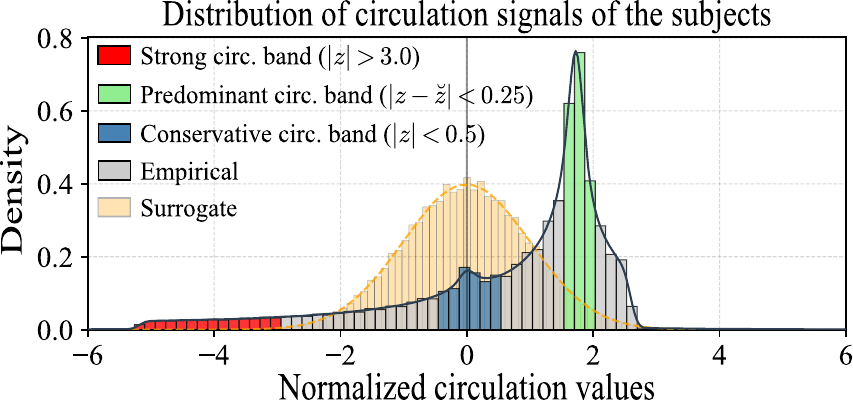}
\label{subfig:curl_distribution}
}
\vspace{-0.1cm}
\\
\subfigure[]
{
\includegraphics[width=0.8\columnwidth,height=3.0cm]{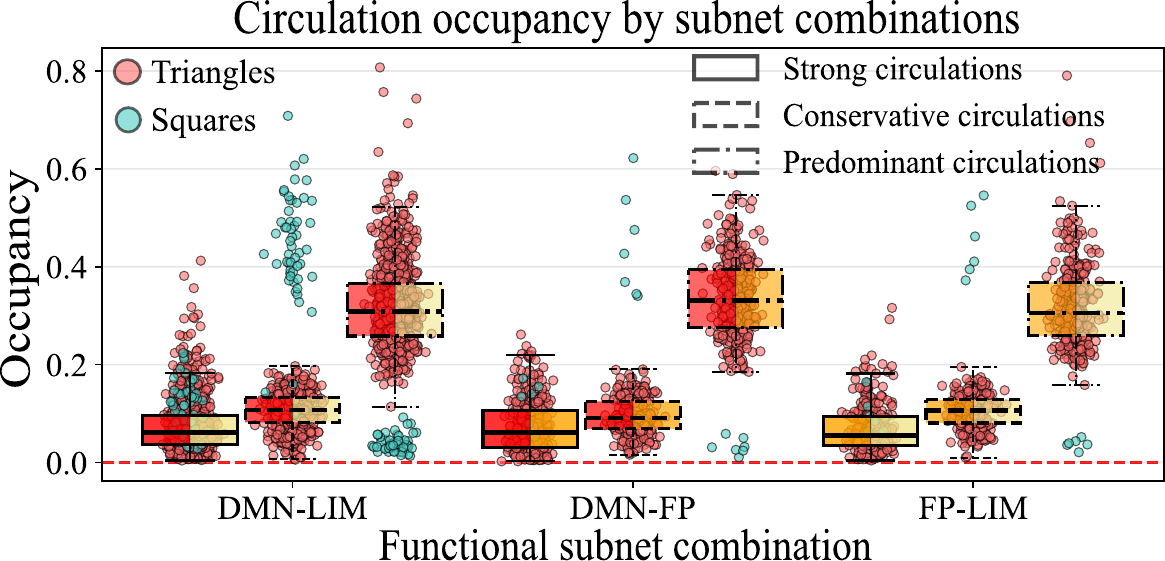}
\label{subfig:occupancy_curl_patterns}
}
\vspace{-0.1cm}
\\
\subfigure[]
{
\includegraphics[width=0.8\columnwidth,height=3.0cm]{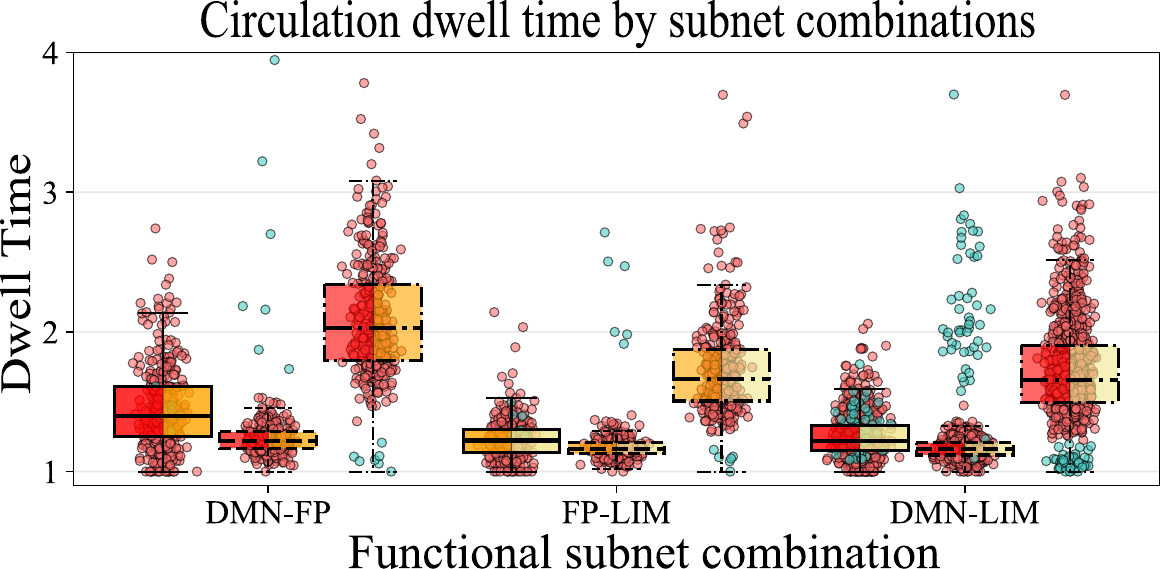}
\label{subfig:dwell_time_curl_patterns}
}
\vspace{-0.1cm}
\\
\subfigure[]
{
\includegraphics[width=0.46\columnwidth,height=2.5cm]{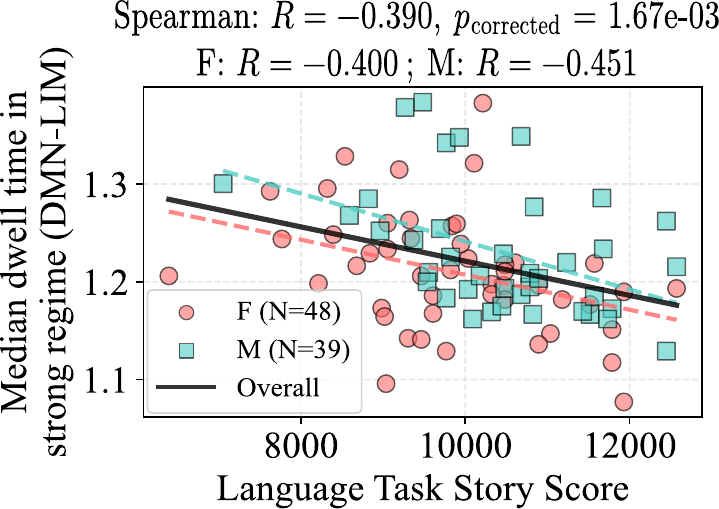}
\label{subfig:correlation_plot_curl_outliers_dmn_lim_language_task_story}
}
\subfigure[]
{
\includegraphics[width=0.46\columnwidth,height=2.5cm]{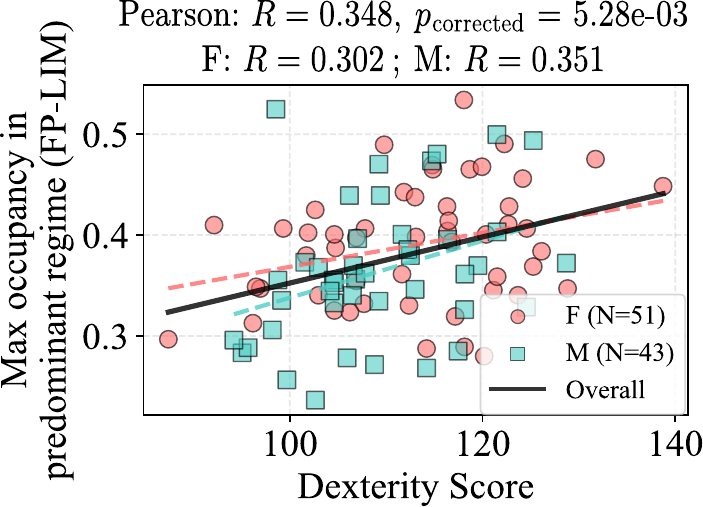}
\label{subfig:correlation_plot_curl_popular_lim_fp_dexterity}
}
\vspace{-0.1cm}
\\
\subfigure[]
{
\includegraphics[width=0.46\columnwidth,height=2.5cm]{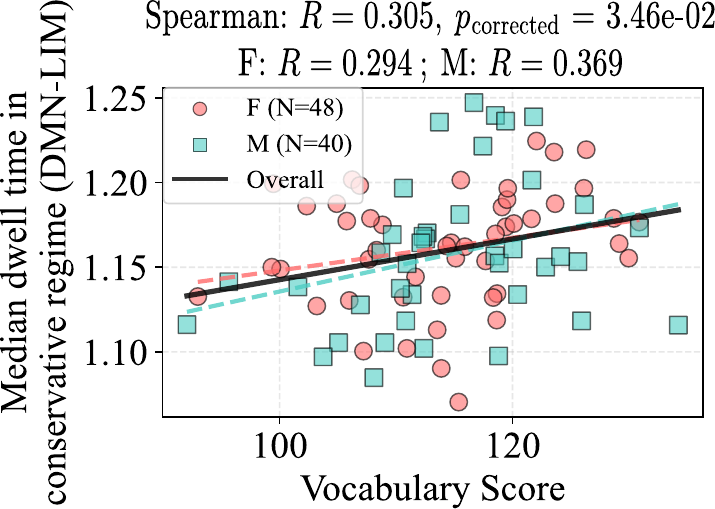}
\label{subfig:correlation_plot_curl_zero_dmn_lim_vocab}
}
\caption{Mesoscale rotational regimes and behavioral associations. \subref{subfig:curl_distribution} Empirical and surrogate z-scored curl distributions across subjects, highlighting conservative, predominant, and strong regimes. \subref{subfig:occupancy_curl_patterns} Fractional occupancy and \subref{subfig:dwell_time_curl_patterns} dwell time across functional-system pairings. Brain–behavior associations between \subref{subfig:correlation_plot_curl_outliers_dmn_lim_language_task_story} DMN–LIM strong-regime dwell time and Language Task Story performance, \subref{subfig:correlation_plot_curl_popular_lim_fp_dexterity} FP–LIM predominant-regime occupancy and motor dexterity, and \subref{subfig:correlation_plot_curl_zero_dmn_lim_vocab} DMN–LIM conservative-regime dwell time and receptive vocabulary. 
}
\label{fig:clinical_curl}
\vspace{-0.6cm}
\end{figure}

{The empirical z-scored circulation distribution across subjects (Fig.~\ref{subfig:curl_distribution}) exhibits a clear multimodal structure and departs from the approximately Gaussian surrogate profile, indicating structured higher-order cyclic coordination beyond stochastic fluctuations. Guided by this finding, we defined three regimes: \emph{strong circulation} ($|z|>3$), capturing transient high-amplitude rotational motifs; \emph{approximately conservative} ($|z|<0.5$), reflecting (near-)zero circulation; and the \emph{predominant circulation} regime centered at the modal value ($|z-\breve{z}|<0.25$), suggesting a stereotyped rotational scale preferentially expressed during rest.}

To quantify regime temporal organization, we measure common dynamic functional connectivity (dFC) metrics~\cite{Cabral2017,leida}, namely, \emph{fractional occupancy}, defined as the fraction of time spent in each state, and \emph{dwell time}, defined as the mean duration of consecutive epochs assigned to the same state. For fair mesoscale comparison, we restrict to 2-cells with FS combinations common across all subjects. Figs.~\ref{subfig:occupancy_curl_patterns} and~\ref{subfig:dwell_time_curl_patterns} depict colored boxplots summarizing the distributions of fractional occupancy and dwell time, stratified by FS combinations and circulation regimes, computed from the empirical edge signals. We also computed the corresponding surrogate-derived distributions, which differed markedly from the empirical ones with $p<.001$.

{Fractional occupancy showed that rotational regimes are not uniformly expressed across subsystem interfaces (Fig.~\ref{subfig:occupancy_curl_patterns}). Interfaces involving transmodal systems, especially DMN, FP, and LIM, exhibited elevated engagement across regimes, compatible with the central role of transmodal/paralimbic regions in large-scale organization \cite{van2013network, Margulies2016Gradient, lim_dmn_hubs}. For DMN-LIM, DMN-FP, and FP-LIM pairings, the predominant regime had the highest occupancy, supporting a shared “default” rotational scale (Fig.~\ref{subfig:curl_distribution}). Cell cardinality further modulated rotation, as square 2-cells showed higher conservative occupancy and lower predominant engagement, suggesting that higher-cardinality motifs favor weakly rotational configurations. Squares were also more prevalent at the DMN-LIM interface, indicating preferential recruitment of higher-order mesoscale interactions in this coupling. Dwell-time analysis revealed metastable interface differences (Fig.~\ref{subfig:dwell_time_curl_patterns}), with DMN-FP pairings showing prolonged dwell in both strong and predominant regimes, indicating that transmodal rotational motifs persist longer once engaged, aligned with findings that core functional networks remain organized during rest \cite{stephen}.}

{We next relate the persistence of subject-specific rotational regimes to HCP behavioral measures~\cite{Barch2013}, focusing on the DMN-LIM and FP-LIM interfaces as candidate loci linking cyclic coordination to cognition.
We observed a significant negative association between the median dwell time of strong DMN-LIM regimes and Language Task Story performance ($R=-0.39$; Fig.~\ref{subfig:correlation_plot_curl_outliers_dmn_lim_language_task_story}).
Because narrative comprehension engages the DMN for semantic integration and contextual updating \cite{menon2023dmn} and limbic-paralimbic circuits for affective salience and motivation \cite{Pessoa2017}, prolonged dwell time in extreme rotational regimes may indicate persistent high-amplitude cyclic coordination. This pattern could be associated with reduced dynamical flexibility and less efficient adaptive reconfiguration during narrative processing.
Maximum fractional occupancy of the predominant rotational regime at the FP-LIM interface was positively associated with motor dexterity ($R=0.348$; Fig.~\ref{subfig:correlation_plot_curl_popular_lim_fp_dexterity}), indicating that greater engagement of this cyclic coordination scale relates to better fine motor performance. Since dexterity requires precise execution alongside adaptive control and sustained engagement, this association aligns with the FP network’s supramodal control role \cite{Niendam2012} and limbic-paralimbic contributions to motivational salience and effort \cite{Pessoa2017}, suggesting that efficient motor behavior depends on stable integration between executive control and motivational systems beyond somatomotor circuitry alone.
Finally, longer persistence in the near-conservative regime at the DMN–LIM interface was positively associated with receptive vocabulary ($R=0.305$; Fig.~\ref{subfig:correlation_plot_curl_zero_dmn_lim_vocab}). As receptive vocabulary relies on stable semantic access and context integration, and DMN with limbic-paralimbic regions (anterior temporal/orbitofrontal) forms an extended default-network architecture interfacing with semantic systems \cite{lim_dmn}, longer dwell in near-zero circulation may reflect a stable baseline coordination mode supporting semantic processing with minimal cyclic reconfiguration.}

\vspace{-0.1cm}
\section{Conclusion}\label{sec:conclusion}
 
In this work, we introduced a new perspective for brain network analysis through a multimodal TSP-based framework, enabling the characterization of higher-order resting-state functional organization under subject-specific anatomical constraints.
For each subject, we learned a 2-cell complex in which statistically significant structural edges define an admissible backbone and HOIs were inferred from time-resolved edge dynamics. This representation enables a Hodge-theoretic treatment of functional interactions as discrete edge flows, separating source-sink redistribution  from higher-order cyclic coordination.

Across 100 healthy HCP young adults, higher-order topology showed marked scale dependence, as node-based HOCs were highly individualized, with low prevalence and near-zero overlap at higher orders, whereas robust mesoscale organization emerged after aggregation by FS combinations. Furthermore, Hodge-based analysis revealed two complementary dynamical modes: a distributed, DMN-centered gradient backbone linking multiple FSs, consistent with large-scale integrative organization; and limbic-centered rotational flows involving primarily DMN, SC, and VIS systems, reflecting recurrent cyclic coordination supported by higher-order motifs. Operator-based analyses showed significant divergence polarization (VIS/VA/LIM biased positive; DMN biased negative) and low-magnitude, low-dispersion SC divergence, consistent with a balanced mediating role. Furthermore, curl values exhibited a structured multimodal distribution, motivating circulation regimes summarized by occupancy and dwell-time statistics. 
Furthermore, exploratory analyses revealed significant associations between divergence- and curl-derived signatures and cognitive-behavioral variability, indicating that curl-supported higher-order coordination complements gradient-driven organization. To our knowledge, no prior work has demonstrated significant brain-behavior associations from topological flow signatures.

Overall, the proposed framework goes beyond pairwise connectomics by establishing cell complexes as a principled topological domain for brain network analysis, enabling polygonal HOI representations not explicitly available in graph-based models. In contrast to prior Hodge-based descriptors on thresholded functional connectivity graphs, our pipeline learns the topology that best represents sparse signals from multimodal data, grounding the higher-order scaffold in structural admissibility and functional edge dynamics. Moreover, unlike existing edge-centric or hypergraph approaches, our framework equips brain signal processing with a physically interpretable discrete calculus, making quantities such as gradient, divergence, and curl directly measurable.
Together, these advances provide a principled TSP-based framework for structure–function modeling and brain-state characterization, while offering a basis to investigate whether higher-order topological fingerprints may prove useful for decoding, stratification, and biomarker-related studies in clinical and cognitive neuroscience.
 
\bibliographystyle{IEEEtran}
\vspace{-0.1cm}
\bibliography{references}

\end{document}